\documentclass[prx, preprint, superscriptaddress, endfloats*]{revtex4-2}
\usepackage{amsmath,hyperref,cleveref,amssymb,amsfonts,graphicx,multirow,color,bm,textcomp,mathtools}
\usepackage{paralist}
\usepackage{srcltx}
\usepackage[all,cmtip]{xy}
\usepackage{mathrsfs}
\usepackage{srcltx,soul,subfigure}
\usepackage{threeparttable,dsfont,bbm}
\usepackage{color}
\usepackage{comment}
\usepackage{dcolumn}

\begin{document}
\title{Complex Magnetic Phases and Unconventional Skyrmion Thermodynamics in Magnetically Intercalated Cr$_{1+x}$Te$_2$ Compounds}
\author{Clayton Conner}
\affiliation {Department of Physics and Astronomy, University of Missouri, Columbia, Missouri 65211, USA}
\author{Santosh Karki Chhetri}
\affiliation {Department of Physics, University of Arkansas, Fayetteville, Arkansas 72701, USA}
\author{Avinash Sah}
\affiliation {Department of Physics and Astronomy, University of Missouri, Columbia, Missouri 65211, USA}
\author{Manoj Gadtoula}
\affiliation {Department of Physics and Astronomy, University of Missouri, Columbia, Missouri 65211, USA}
\author{Dilan M. Gamachchi}
\affiliation {Department of Physics and Astronomy, University of Missouri, Columbia, Missouri 65211, USA}
\author{Indeewari M. Karunarathne}
\affiliation {Department of Physics and Astronomy, University of Missouri, Columbia, Missouri 65211, USA}
\author{Steven Kelley}
\affiliation {Department of Chemistry, University of Missouri, Columbia, Missouri 65211, USA}
\author{Andrew C. Meng}
\affiliation {Department of Physics and Astronomy, University of Missouri, Columbia, Missouri 65211, USA}
\affiliation {MU Materials Science \& Engineering Institute, University of Missouri, Columbia, Missouri 65211, USA}
\author{Jacob Cook}
\affiliation {Department of Physics and Astronomy, University of Missouri, Columbia, Missouri 65211, USA}
\affiliation {Materials Sciences and Technology Division, Oak Ridge National Laboratory, Oak Ridge, Tennessee 37830, USA}
\author{Cheng Zhang}
\affiliation {Department of Physics and Astronomy, University of Missouri, Columbia, Missouri 65211, USA}
\author{Jin Hu}
\affiliation {Department of Physics, University of Arkansas, Fayetteville, Arkansas 72701, USA}
\affiliation {Arkansas Materials Institute, University of Arkansas, Fayetteville, Arkansas 72701, USA}
\affiliation {MonArk NSF Quantum Foundry, University of Arkansas, Fayetteville, Arkansas, 72701, USA}
\affiliation {Smart Ferroic Materials Center, University of Arkansas, Fayetteville, Arkansas, 72701, USA}
\author{Yue Li}
\affiliation {Materials Science Division, Argonne National Laboratory, Lemont, Illinois 60439, USA}
\author{Hoyeon Jeon}
\affiliation {Center for Nanophase Materials Sciences, Oak Ridge National Laboratory, Oak Ridge, TN 37831, USA}
\author{An-Ping Li}
\affiliation {Center for Nanophase Materials Sciences, Oak Ridge National Laboratory, Oak Ridge, TN 37831, USA}
\author{Zheng Gai}
\affiliation {Center for Nanophase Materials Sciences, Oak Ridge National Laboratory, Oak Ridge, TN 37831, USA}
\author{Guang Bian}\email{biang@missouri.edu}
\affiliation {Department of Physics and Astronomy, University of Missouri, Columbia, Missouri 65211, USA}
\affiliation {MU Materials Science \& Engineering Institute, University of Missouri, Columbia, Missouri 65211, USA}

\newpage
\begin{abstract}
{Chemical intercalation provides a powerful route for tuning the electronic and magnetic properties of transition metal dichalcogenides (TMDs), enabling the emergence of magnetic phases and phenomena that are otherwise inaccessible in their pristine forms. Here, we investigate the structural, electronic, and magnetic properties of magnetically intercalated Cr$_{1+x}$Te$_2$ through a combination of scanning tunneling microscopy, angle-resolved photoemission spectroscopy, magnetization, electrical transport, and Lorentz transmission electron microscopy measurements. We uncover a rich magnetic phase diagram comprising of paramagnetic, in-plane spin-fluctuation, out-of-plane ferromagnetic, and field-induced Néel-type skyrmion phases. In particular, Lorentz transmission electron microscopy directly reveals the formation of Néel-type skyrmion textures slightly below the Curie temperature, with their configurations evolving strongly with temperature and magnetic field. The skyrmions exhibit pronounced thermal hysteresis and coexist with stripe domains, indicative of a complex magnetic energy landscape arising from the competition among disorder-induced local Dzyaloshinskii–Moriya interactions, perpendicular magnetic anisotropy, and dipolar interactions. Our results demonstrate that chemical intercalation can profoundly reshape the magnetic interactions and phase behavior of TMDs, establishing Cr$_{1+x}$Te$_2$ as a versatile platform for exploring tunable magnetism, skyrmion physics, and unconventional thermodynamic phenomena in low-dimensional quantum materials.}

\end{abstract}


\maketitle

\newpage

\section{Introduction}

Van der Waals (vdW) materials continue to attract significant interest due to their diverse physical properties, potential practical device applications, and relative ease of fabrication. Among these systems, transition metal dichalcogenides (TMDs) are the most studied vdW materials on account of the  multitude of exciting physical features found in these materials, ranging from superconductivity, topological phenomena, two-dimensional magnetism, and spintronics applications \cite{doi:10.1126/science.aac9439,doi:10.1126/science.aav4450}. Several layered materials such as Fe$_3$GeTe$_2$ \cite{PhysRevB.93.134407,PhysRevB.93.144404}, Cr$_2$Ge$_2$Te$_6$  \cite{Gong2017-iw}, and CrI$_3$ \cite{Huang2017-vl} have been intensively investigated due to their 2D magnetic properties and their potential for spin-based device applications.  Among these layered magnetic materials, CrTe$_2$ stands out for its robust magnetic ordering in the two-dimensional limit and its exceptional tunability of magnetic properties via chemical doping \cite{Zhang2021-cm, zhang2025advances}. In addition to the intrinsic properties of CrTe$_2$, a family of related compounds can be realized through self-intercalating excess Cr atoms within the vdW gaps, yielding materials such as Cr$_2$Te$_3$, Cr$_3$Te$_4$, and Cr$_5$Te$_8$. These compounds, commonly described by the formula Cr$_{1+x}$Te$_2$, preserve the layered nature of the parent CrTe$_2$ structure while exhibiting distinct structural, magnetic, and electrical properties \cite{JDijkstra_1989}. Depending on the concentration of Cr intercalation, these materials crystallize in trigonal or monoclinic phases, exhibit tunable Curie temperatures ($T_\mathrm{C}$), and host a rich variety of magnetic states \cite{zhang2025advances, PhysRevMaterials.4.114001}. 

Magnetically intercalated Cr$_{1+x}$Te$_2$ compounds provide a promising platform for exploring emergent chiral spin textures, such as magnetic skyrmions. Skyrmions are topologically nontrivial quasiparticles with complex spin configurations confined to regions typically spanning tens to hundreds of nanometers in radius \cite{doi:10.1021/acs.chemrev.0c00297}. The spin structure of skyrmions generally forms a chiral loop where the central moment points in the opposite direction of the edge moments. Due to their topological nature, they remain robust to external perturbations such as electric and magnetic fields, making them a very exciting prospect for spintronics applications \cite{10.1063/5.0072735}. In prototypical chiral magnets, skyrmions are realized by a balance of symmetric exchange interaction and chiral exchange interaction known as the Dzyaloshinskii-Moriya Interaction (DMI), which arises in the presence of strong spin-orbit coupling (SOC) and broken inversion symmetry \cite{DZYALOSHINSKY1958241,PhysRev.120.91}. The DMI gives rise to canting of the spins and ultimately enables the chiral spin texture found in skyrmions. The DMI can stabilize Bloch type skyrmions in bulk noncentrosymmetric crystals, but it commonly favors Néel-type skyrmions in thin films due to the interfacial symmetry breaking. More recently, increasing experimental and theoretical evidence has demonstrated that chiral spin textures can also emerge in centrosymmetric materials without a global bulk DMI, indicating that multiple magnetic interactions may cooperate to stabilize skyrmion-like states \cite{doi:10.1021/acs.nanolett.9b03453,PhysRevLett.129.137202}. In chemically intercalated magnetic vdW materials, partial site occupancy, vacancies, and local lattice distortions can locally break inversion symmetry over nanometer length scales, generating spatially inhomogeneous DMI even when the average crystal structure remains centrosymmetric. Although these local chiral interactions average to zero over the crystal, they can cause internal spin rotation of magnetic textures and stabilize non-collinear spin textures such as skyrmions \cite{doi:10.1021/acs.nanolett.4c01486, PhysRevB.111.054417, PhysRevB.93.184413, doi.org/10.1002/adma.202108637, Saha2022}. In addition, other spin interactions such as exchange interactions, Ruderman-Kittel-Kasuya-Yosida (RKKY) coupling in itinerant magnets, magnetic anisotropy, and long-range dipole-dipole interactions can collectively affect non-collinear spin textures of skyrmions and thus determine the skyrmion stability and characteristic length scale \cite{PhysRevB.93.184413,PhysRevB.93.064430,PhysRevLett.124.207201, PhysRevB.97.014425}. In particular, despite the central roles of perpendicular magnetic anisotropy (PMA) and long-range dipolar interactions in determining the domain morphology, their relative contributions to the formation of skyrmions remain challenging to disentangle experimentally. Establishing a quantitative link between these competing energy scales and the observed magnetic textures is therefore essential for understanding skyrmion formation in disordered centrosymmetric materials.

In this work, we present a comprehensive study of the structural, electrical, and magnetic properties of single-crystal Cr$_{1+x}$Te$_2$ ($x=$ 0.205). Single-crystal X-ray diffraction measurements confirm that this compound crystallizes in a globally centrosymmetric structure with partially occupied intercalated Cr sites, which introduce an intrinsic source of structural disorder into the lattice. Magnetization, transport, and AC susceptibility measurements reveal a rich magnetic phase diagram characterized by multiple anomalies that correspond to a ferromagnetic state, magnetic skyrmions, and an in-plane spin-fluctuation state. These observations demonstrate that the magnetic ground state of Cr$_{1+x}$Te$_2$ is governed by a complex energy landscape containing several competing magnetic phases rather than a single homogeneous ferromagnetic state. In addition, our Lorentz transmission electron microscopy (LTEM) measurements reveal the formation of Néel-type skyrmions below the Curie temperature $T_\mathrm{C}$. Combined with the intrinsic Cr-site disorders revealed by single-crystal X-ray diffraction (SCXRD), these observations suggest that local symmetry breaking and spatially inhomogeneous chiral interactions give rise to the magnetic skyrmions. Our results further demonstrate the evolution of the skyrmion phase over a wide range of temperatures and applied magnetic fields and reveal unconventional skyrmion thermodynamics in which the size and density of skyrmions are governed by a competition between perpendicular magnetic anisotropy and long-range dipolar interactions.

\section{Results and Discussion}
Cr$_{1+x}$Te$_2$ single crystals were synthesized using the self-flux method. The structure of these crystals was measured using SCXRD, with further confirmation by scanning tunneling microscope (STM), transmission electron microscope (TEM), and scanning transmission electron microscope (STEM). Figure~1(a,b) shows the crystal structure as measured from SCXRD. The quasi-2D structure that is characteristic of these materials is evident in Fig.~1(a), with layers of CrTe$_2$ and excess Cr atoms intercalated within the vdW gaps that exist in the parent compound of CrTe$_2$. The excess Cr atoms occupy two crystallographically distinct intercalation sites, denoted Cr$_3$ and Cr$_4$, both of which are only partially occupied. These nonequivalent statistical occupancies (0.430 and 0.194, respectively) introduce intrinsic structural disorder throughout the lattice while preserving the average crystallographic symmetry. With this, the compound has a precise stoichiometry (within $\pm 0.05$) of Cr$_{4.82}$Te$_8$ ($x$ = 0.205 in Cr$_{1+x}$Te$_2$), which is slightly Cr-poor compared to the stoichiometric Cr$_5$Te$_8$ ($x$ = 0.25). This compound crystallizes into the monoclinic space group C2/m with SCXRD-measured lattice parameters of $a = 11.982(1)$~\AA, $b = 7.8109(7)$~\AA, $c = 9.0316(8)$~\AA, $\alpha=\gamma=90^\circ$, and $\beta=131.542(3)^\circ$. The monoclinic distortion arises from the non-orthogonal angle between the $a$- and $c$-axes, while the average global inversion symmetry is preserved within the C2/m structure. Using a Cartesian basis with $\mathbf{a} \parallel \hat{z}$ and $\mathbf{b} \parallel \hat{y}$, the monoclinic lattice vector $\mathbf{c}$ lies in the $xz$-plane and can be written as $\mathbf{c}=(c \sin\beta)\hat{x}+(c \cos\beta)\hat{z}$. Compared to the pristine hexagonal lattice of CrTe$_2$ ($a'=3.8$~\AA, $c'=6.1$~\AA), the monoclinic structure of Cr$_{1+x}$Te$_2$ exhibits an expansion of the in-plane periodicity, with $b \approx 2a'$. Additionally, the system shows a contraction along the CrTe$_2$ stacking direction as $a < 2c'$. These structural distortions can be attributed to the partially occupied Cr$_3$ and Cr$_4$ intercalation sites, which are expected to generate a spatially inhomogeneous local lattice environment. Although the average crystal structure remains centrosymmetric, the intrinsic site disorder provides a natural source of local inversion symmetry breaking and chiral spin interactions that can facilitate the formation of magnetic skyrmions. 

The crystal structure of this compound was further verified by STM and TEM, as shown in Figs.~1(c-e). The high-angle annular dark field (HAADF) STEM image (Fig.~1(c)) and selected area electron diffraction (SAED) (Fig.~1(d)) of Cr$_{1+x}$Te$_2$ confirm a high structural quality consistent with the SCXRD result. The diffraction pattern is directly indexed using the SCXRD structure. A corresponding simulated SAED pattern is provided in the Supplementary Materials. The pattern is acquired along the [100] zone axis, such that the observed reflections correspond to the reciprocal lattice of the $b$-$c$ plane. Despite the monoclinic symmetry of the bulk structure, the SAED pattern exhibits a pronounced hexagonal pseudo-symmetry of the diffraction spots. This arises from the projection of the three-dimensional lattice along the $a$ axis, where the Te sublattice within the $b$-$c$ plane forms a nearly trigonal network. Due to the larger atomic scattering factor of Te relative to Cr, the structure factor is dominated by the Te sublattice  \cite{Sarney2003-um}. This results in the approximately hexagonal arrangement of diffraction spots, despite the lower symmetry of the monoclinic unit cell. The atomic-resolution STM image of an $in~situ$ cleaved Cr$_{1+x}$Te$_2$ surface can be found in Fig.~1(e), revealing a trigonal periodicity consistent with the symmetry observed in SAED results. The apparent trigonal symmetry of the STM result can be understood by the surface sensitivity of the STM itself, which probes the topmost layer of the crystal. In this material, the cleaved surface corresponds to a CrTe$_2$ terminated layer, which intrinsically possesses trigonal periodicity. As a consequence, both STM and SAED emphasize the same in-plane pseudo-hexagonal symmetry. The in-plane lattice constant is measured to be 4.05~\AA~by fast Fourier transform (FFT) of the atomic resolution STM image, shown in the inset of Fig.~1(e). This value corresponds to half of the $b$ lattice constant obtained from the SCXRD-measured unit cell, indicating good agreement between surface-sensitive and bulk measurements. Additional topographic STM measurements confirm a step height consistent with approximately one-half of the SCXRD-measured $a$ lattice constant (see Supplementary Materials). The coexistence of a well-defined long-range crystal structure and intrinsically disordered intercalated Cr sublattice plays an important role in the magnetic properties of Cr$_{1+x}$Te$_2$. Because the exchange interactions in Cr$_{1+x}$Te$_2$ compounds are highly sensitive to Cr occupancy and local coordination, the statistical distribution of Cr$_3$ and Cr$_4$ sites gives rise to spatial variations in the local exchange coupling and magnetic anisotropy. Such local variations induced by Cr intercalants play a pivotal role in shaping the complex magnetic phase diagram, as discussed below.

The electronic structure of Cr$_{1+x}$Te$_2$ was characterized using angle-resolved photoemission spectroscopy (ARPES). Figure 1(f) shows the measured band structure of Cr$_{1+x}$Te$_2$ near the Fermi level along the M-$\Gamma$-M direction. There are two hole pockets clearly visible from the spectrum with one crossing the Fermi energy, giving rise to the metallic character of  Cr$_{1+x}$Te$_2$. Notably, these ARPES features closely resemble previously reported band structures of CrTe$_2$ \cite{Zhang2021-cm}. This similarity suggests that this Cr$_{1+x}$Te$_2$ system can be viewed approximately as a  CrTe$_2$ system that has been doped with excess Cr atoms. To further support this interpretation, simulated ARPES spectra for CrTe$_2$ are presented in the Supplementary Materials and show good agreement with the experimental data, except for a shift in the Fermi energy. In layered vdW materials, chemical intercalation typically preserves the strong in-plane bonding of the host lattice, allowing the electronic structure to be manipulated without fundamentally altering the underlying band dispersion  \cite{LI2024118}. The excess Cr intercalated in Cr$_{1+x}$Te$_2$ shifts the Fermi energy relative to CrTe$_2$, as demonstrated by the consistency between experimental and simulated ARPES results. While deviations from the band structure of parent compounds are allowed by hybridization between intercalants and host layers \cite{Wang2020-lu}, the overall similarity of the measured band structure indicates that the CrTe$_2$-derived bands dominate the low-energy electronic structure of Cr$_{1+x}$Te$_2$.

Having established the structural and electronic properties of this compound, we next investigate the spin and magnetic behavior of Cr$_{1+x}$Te$_2$ ($x=$ 0.205) through field- and temperature-dependent magnetization measurements. The field-dependent magnetization $M(H)$ reveals a pronounced out-of-plane (OOP) easy axis. This is confirmed by the rapid saturation of the magnetic moment under an out-of-plane (OOP) applied field, whereas the in-plane (IP) magnetization does not saturate even at the applied fields of 4~T (Figs.~2(a,b)). This behavior is indicative of a perpendicular magnetic anisotropy (PMA) in the system, which is consistent with the previous results of Cr$_{1+x}$Te$_2$ materials \cite{doi:10.1021/acsnano.0c05534,Zhang2021-cm, Huang_2021}. The OOP magnetic moment saturates to approximately 1.9 $\mu_B$ per Cr atom under a magnetic field of 0.35~T. The coercive field was determined to be approximately 300~Oe, indicating relatively soft magnetic behavior (see Fig.S4 in Supplementary Materials). The measured value of the magnetic moment is less than the theoretically predicted value of 3.0 $\mu_B$ in an ionic model of the system. The difference suggests that the magnetic moments in the system are not fully localized, which is consistent with itinerant magnetism arising from partially delocalized Cr 3$d$ electrons. This is supported by the ARPES results, which confirm a metallic electronic structure, a necessary condition for itinerant magnetism. 

Temperature-dependent magnetization $M(T)$ measurements were performed under an applied field of 500~Oe in the IP (Fig.~S5 in Supplementary Materials) and OOP (Fig.~2(c)) directions. A clear ferromagnetic transition was observed at the Curie temperature, $T_\mathrm{C}$ = 226~K. The in-plane magnetization remains substantially smaller than the out-of-plane response across the measured temperature range. This is consistent with the $M(H)$ results and further confirms the perpendicular easy axis. In the OOP measurement, a pronounced bifurcation emerges between the zero-field-cooled (ZFC, sample was cooled down at zero field, and measured under field during warming up) and field-cooled (FC, sample was measured under field during cooling down) curves below $T_\mathrm{C}$. Both the ZFC and FC curves were taken under an applied field of $H=500$~Oe, with $H\parallel z$. As the temperature decreases, the FC magnetization increases gradually while the ZFC magnetization progressively decreases and approaches nearly zero. The splitting of the ZFC and FC curves arises from the FM domain effect \cite{Yi_2017}, as the structural disorder from the partially occupied intercalated Cr sites can facilitate domain wall pinning. However, the ZFC measurement was done at an applied field that exceeds the coercive field of 300~Oe, meaning that the suppressed ZFC response cannot be solely attributed to the domain effect. Instead, the persistence of a nearly compensated magnetic state under applied field suggests the existence of metastable and non-uniform spin configurations. This bifurcation effect becomes pronounced below approximately 170~K. The onset of this bifurcation coincides with the temperature range in which LTEM measurements reveal the formation of magnetic skyrmions, pointing toward a magnetic regime where competing interactions and non-collinear spin structures play a crucial role. Once these skyrmion states are formed during zero-field cooling, they require fields larger than 500~Oe to fully polarize, thereby producing the observed bifurcation. 

The magnetic nature of this material can be further probed by transport and AC susceptibility measurements. Figure 2(d) presents the longitudinal resistivity, $\rho_{xx}(T)$. A pronounced change in slope occurs at the Curie temperature of 226 K, consistent with the $T_\mathrm{C}$ measured from M(T) measurements. This rapid decrease in $\rho_{xx}$ below the $T_\mathrm{C}$ is a well-known feature found in itinerant ferromagnets. When the system is above the Curie temperature, magnetic moments fluctuate, resulting in a large contribution of spin scattering to the resistivity. As the system transitions to the ferromagnetic phase, long-range ordering of the moments develops, and this spin scattering effect decreases rapidly. In addition to the slope change at $T_\mathrm{C}$, a very subtle kink occurs in the $\rho_{xx}(T)$ curve near 172~K, which is highlighted by the inset in Fig.~2(d). This feature appears at the onset temperature of the ZFC–FC bifurcation, indicating its close association with the stabilization of magnetic skyrmions.

AC magnetic susceptibility measurements were performed to provide insight into the complex nature of the magnetic ordering in this system. Figure~2(e) shows the results of this measurement done at a frequency of 100 Hz. A comprehensive magnetic phase diagram can be extracted from the real part of $\chi'$ and imaginary part $\chi''$ of AC magnetic susceptibility for both in-plane and out-of-plane configurations, presented in Figure~2(f). As the system is cooled from room temperature, a weak knee-like feature emerges in the IP $\chi'$ curve at 247~K, while no corresponding feature is observed in the out-of-plane (OOP) $\chi'$ curve. This magnetization behavior can be attributed to the development of in-plane spin fluctuations (IPSF), consistent with previous reports on other Cr$_{1+x}$Te$_2$ compounds \cite{https://doi.org/10.1002/aelm.202500599,Bigi2025}. The $2\times 2$ superlattice structure brings about an effective in-plane antiferromagnetic exchange between the intercalated atoms and the CrTe$_2$ layers according to the DFT calculations \cite{Bigi2025}. The in-plane spin component of the intercalated Cr develops three equivalent energy minima. Consequently, the spin of intercalated Cr tilts toward one of the three energetically favored directions, leading to the in-plane spin fluctuations. These results indicate a correlation between the $2\times2$ superstructure and the exchange coupling between the Cr intercalants and the Cr atoms in the host lattice, which precedes the establishment of long-range ferromagnetic order. At 226~K, the ferromagnetic transition produces prominent peaks in both $\chi'$ and $\chi''$. Notably, an additional feature emerges at approximately 172~K in the OOP configuration, visible in both $\chi'$ and $\chi''$. The concurrence of both reactive ($\chi'$) and dissipative ($\chi''$) components at this temperature suggests a reorganization of the magnetic texture. This feature coincides with the observed resistivity kink and implies a modification of spin-dependent transport, consistent with a transition from a uniform ferromagnetic state to a nonuniform or modulated magnetic texture. The sharpness of this anomaly further supports the presence of a distinct phase boundary rather than a gradual crossover. This interpretation is corroborated by both transport and LTEM measurements. The 172 K susceptibility anomaly coincides with the resistivity kink and lies within the temperature window where LTEM directly images Néel skyrmion bubbles. Together, these observations indicate that the low-temperature anomaly marks the onset of a distinct skyrmion-bubble regime rather than a simple modification of the ferromagnetic order parameter. The resulting phase diagram, shown in Fig~2(f), thus identifies four successive magnetic phase regimes upon cooling: a paramagnetic state, an in-plane spin-fluctuation regime, an out-of-plane ferromagnetic phase, and finally an out-of-plane ferromagnetic state hosting Néel skyrmion bubbles. This rich magnetic phase diagram highlights the complex energy landscape of Cr$_{1+x}$Te$_2$. It provides a framework for understanding how magnetic anisotropy, dipolar interactions, and structural disorder compete and collectively give rise to the unconventional thermodynamics of magnetic skyrmions.

The concurrence of the OOP AC susceptibility anomaly and the resistivity kink within the temperature range of ZFC/FC bifurcation suggests the development of a non-uniform magnetic spin texture. However, bulk probes alone cannot distinguish between domain effects, spin reorientation, or topologically nontrivial spin configurations. To directly visualize the magnetic texture in real space, LTEM measurements were systematically performed.  Figure~3(a) presents a schematic of a Néel-type magnetic skyrmion, characterized by the radial rotation of spins outward from the core. This spin texture produces a distinct contrast in LTEM that appears only when the sample is tilted relative to the electron beam. At zero tilt angle, the radial spin texture produces no beam deflection, resulting in minimal visible contrast. An in-plane projection of the magnetization develops when the sample is tilted by an angle $\alpha$ relative to the beam, generating an asymmetric bright-dark contrast \cite{McVitie2018-iw}. A thin lamella of Cr$_{1+x}$Te$_2$ was prepared using focused ion-beam (FIB) milling in the (100) orientation for LTEM measurements. The lamella was first cooled from 300~K to 120~K under zero applied field to check initial spin textures. When cooled without an applied field, the system only exhibits stripe-like domains. The Néel-type character of the domains is confirmed in Fig.~3(b), with no magnetic contrast when $\alpha$ is zero. Once the lamella is tilted in the positive or negative direction, the Néel-type domains appear with contrast reversal from negative to positive angle $\alpha$. An applied magnetic field parallel to the easy axis during cooling of the lamella results in the formation of magnetic skyrmions. The appearance of skyrmions occurs around 215 K, close to the measured $T_\mathrm{C}$, as shown in Fig.~3(c) with a Néel-type contrast. 


The temperature evolution of the magnetic texture provides key insight into the stability of the skyrmion states. Upon cooling the sample, skyrmions first emerge near the Curie temperature under a field of 300~Oe. They are densely populated and possess an average diameter of approximately 180~nm. In addition, Néel-type domain walls,  which were seen at zero field, remain present, indicating the persistence of the underlying chiral magnetic texture. Upon further cooling, the skyrmions and stripe domains gradually expand in size, accompanied by a reduction in their density, as illustrated in Fig.~3(c). Notably, the LTEM image taken near 175 K shows the onset of the skyrmion expansion, indicating a modification of the magnetic texture within the skyrmion phase. This temperature coincides with the resistivity kink and the peak observed in the AC susceptibility. Below this temperature, the skyrmions increase in size, indicating a temperature-driven redistribution of magnetic energy that stabilizes fewer but larger spin textures. Upon warming from 100 K, this evolution is not reversible, as shown in Fig. 3(d). The skyrmions remain in their enlarged configuration until they vanish above 220~K.  

Interestingly, the temperature evolution of the skyrmion phase shows a pronounced dependence on thermal history. Initially, the lamella was cooled directly from room temperature, during which skyrmions were observed; however, they vanished upon subsequent warming, suggesting only limited metastability. In contrast, when the lamella was first heated to 350~K - above the Curie temperature reported for CrTe$_2$, the skyrmions displayed strong hysteresis and persisted to significantly higher temperatures. This behavior implies that the stabilization of the skyrmion phase is highly sensitive to magnetic disorder. Heating above 350~K effectively resets local magnetic correlations, defect configurations, or competing magnetic clusters. The observed hysteresis of the skyrmion phase cannot be explained solely by conventional DMI, suggesting that additional mechanisms play a role in stabilizing the skyrmions.

Magnetic field-dependent LTEM measurements were performed at a fixed temperature of 170~K to probe the stability of the skyrmion phase under external magnetic tuning. As shown in Figs. 4(a-d), the system exhibits a coexistence of stripe domains and skyrmion bubbles at low magnetic fields, consistent with a stripe-derived origin of the skyrmions. With increasing applied magnetic field, both the skyrmion diameter and their density decrease continuously. Stripe domains are largely suppressed at 1100~Oe, whereas skyrmion bubbles persist up to $\sim$1600~Oe, beyond which the system becomes fully spin-polarized. After saturation, reducing the magnetic field to zero drives the system back into a stripe-domain-dominated ground state, demonstrating the reversibility of the magnetic phase transition with changes in the applied magnetic field. This behavior is characteristic of dipolar-driven magnetic textures. 
The quantitative evolution of skyrmion size with temperature and magnetic field (Figs.~4(e,f)) reveals a consistent trend: cooling enhances dipolar interactions and promotes skyrmion expansion, while increasing field induces contraction via Zeeman coupling. The coexistence of these effects, together with the observed thermal hysteresis and its correlation with the ZFC/FC bifurcation, indicates that the magnetic state is governed by a complex energy landscape with substantial barriers arising from the structural disorder.


The magnetic textures observed in Cr$_{1+x}$Te$_2$ emerge from a delicate balance of competing interactions within a globally centrosymmetric lattice containing intrinsically disordered Cr intercalants \cite{Saha2022}. The presence of stripe domains at low field and their recovery upon field removal indicate that the zero-field ground state favors a configuration that minimizes magnetostatic energy. The stripe instabilities under applied fields give rise to the skyrmion-like bubbles. This idea can be examined by investigating the perpendicular magnetic anisotropy of Cr$_{1+x}$Te$_2$ and its temperature dependence. The in-plane (hard-axis) magnetization curves indicate a large anisotropy field exceeding 7~T across the measured temperature range, confirming a robust PMA in the system. Despite the large anisotropy scale, the high-field response of the IP magnetization shows only weak temperature dependence between 200~K and 100~K, indicating the intrinsic anisotropy energy, $K_u$, remains constant in this temperature range. Using $K_u \approx \frac{1}{2}\mu_0 M_s H_k$, with $M_s \sim 2.0~\mu_B$ per Cr, converted to volume magnetization using the unit cell volume calculated from the SCXRD lattice parameters, and $\mu_0 H_k \gtrsim$ 7–8 T, yields $K_u \sim 10^6$ J/m$^3$. This places the system in a regime where anisotropy and dipolar energies are comparable, consistent with the formation of stripe and bubble domains \cite{VANDERLAAN2003107,KWON20102742}. In contrast, the saturation magnetization shows a strong temperature dependence below the $T_\mathrm{C}$ of 226~K, increasing from approximately 1.5 $\mu_B$/Cr at 200 K to 1.96 $\mu_B$/Cr at 100 K. Since the dipolar magnetostatic energy scales with $M_s^2$, the increase in saturation magnetization upon cooling enhances the dipolar energy by approximately 70\% over the measured temperature range. Consequently, the effective anisotropy $K_{\mathrm{eff}} = K_u - \frac{1}{2}\mu_0 M_s^2$ is reduced at lower temperatures, shifting the balance of interactions towards dipolar contributions. The coincidence of this evolution with the onset of skyrmion expansion below approximately 175 K suggests that the observed expansion of the magnetic textures is driven by a temperature-dependent redistribution of the magnetic energy landscape.

The resulting evolution of skyrmion size can be understood within a micromagnetic framework, in which the total energy is determined by a competition between domain wall energy and magnetostatic contributions \cite{hubert1998magnetic,Woo2016-dj}. The domain-wall energy scales with perimeter as $E_{\mathrm{wall}} \propto R$, reflecting the cost of spin rotation across the texture, while the magnetostatic energy scales with area as $E_{\mathrm{dip}} \propto M_s^2 R^2$, favoring domain expansion. The equilibrium domain size corresponds to a minimum of this energy balance. As temperature decreases, the increase in $M_s$ strengthens the magnetostatic driving force while the reduction in $K_{\mathrm{eff}}$ reduces the relative stability of uniform domains, together shifting the energy minimum toward larger characteristic radii. The onset of this expansion near 175~K, coincident with the resistivity kink and AC susceptibility peak, indicates a crossover in the magnetic energy landscape rather than a simple thermal evolution. Because the domain size ($R$) is comparable to the lamella thickness ($t \approx 100$~nm), the system lies in a regime where dipolar interactions cannot be adequately described within a strictly two-dimensional approximation. Magnetostatic interactions retain significant thickness dependence, reinforcing their dominant role in setting the equilibrium length scale.


Magnetic field-dependent LTEM measurements provide a complementary probe of this energy balance. At 170~K, the system exhibits a mixture of stripe domains and skyrmion bubbles at low fields, consistent with a stripe-derived origin of the skyrmions. As the applied magnetic field is increased, both the size and density of the skyrmions decrease progressively, with complete polarization occurring near 1600~Oe. Within the same micromagnetic framework, the applied field introduces a Zeeman contribution, $E_\mathrm{Z} \sim -\pi R^2 \mu_0 M_s H$, which favors alignment of the magnetization with the field and penalizes the reversed skyrmion core. As a result, increasing field shifts the energy minimum toward smaller radii and ultimately destabilizes the skyrmion state. Upon reducing the magnetic field to zero, the system relaxes back into a stripe-like ground state, demonstrating the reversibility of the magnetic transition. This field-driven evolution is characteristic of dipolar-stabilized magnetic domains and stands in contrast to the relatively fixed size and robustness typically associated with skyrmions stabilized by bulk DMI interactions. The existence of skyrmions only within a finite temperature window between approximately 100~K and 215~K further supports this interpretation. Near $T_\mathrm{C}$, the reduced magnetization of the system suppresses dipolar interactions, limiting the stability of well-defined bubble domains. Upon cooling, the increasing dipolar energy drives stripe instabilities that fragment into skyrmion-like bubbles under the applied field. However, at lower temperatures the continued growth of dipolar energy favors larger domain structures rather than a higher density of skyrmions, resulting in the experimentally observed increase in skyrmion size and reduction in density. The itinerant character of the magnetism in this compound, together with a reduced ordered moment and softened exchange stiffness, further enhances sensitivity to disorder and facilitates domain coarsening at lower temperatures.


The structural disorder introduced by the partially occupied Cr$_3$ and Cr$_4$ intercalation sites provides a natural framework for understanding the Néel character of the observed magnetic textures. Although the average C2/m crystal structure is globally centrosymmetric and therefore does not support a uniform bulk DMI, the statistical distribution of intercalated Cr atoms is expected to produce local variations in coordination, exchange pathways, and inversion symmetry. Such local symmetry breaking can generate spatially inhomogeneous chiral interactions that average to zero over the crystal while remaining finite on nanometer length scales \cite{doi:10.1021/acsnano.5c08312}. Within this picture, dipolar interactions and perpendicular magnetic anisotropy determine the stability and characteristic size of the bubble domains, whereas local chiral interactions favor Néel-type domain-wall configurations and define the internal spin rotation of the textures \cite{KWON20148}. Taken together, these results demonstrate that the skyrmion-like textures in Cr$_{1+x}$Te$_2$ emerge from a cooperative interplay between disorder, exchange, perpendicular magnetic anisotropy, and long-range dipolar interactions. This places Cr$_{1+x}$Te$_2$ in a regime where both disorder-induced DMI and dipolar interactions play important roles in determining the thermodynamics of Néel-type skyrmions, distinguishing it from conventional bulk DMI-stabilized skyrmion systems.

\section{Conclusion}
In summary, we investigated the structural, electronic, and magnetic properties of Cr$_{1+x}$Te$_2$ through a combination of STM, ARPES, magnetization, transport, AC susceptibility, and LTEM measurements. The results reveal a rich magnetic phase diagram comprising distinct phases, including a paramagnetic phase, an in-plane spin-fluctuation regime, an out-of-plane ferromagnetic phase, and a Néel-type skyrmion phase that emerges under an applied magnetic field. Furthermore, LTEM results directly demonstrate the formation of Néel-type skyrmion textures slightly below the Curie temperature, which evolve strongly with temperature and applied field. These skyrmions exhibit pronounced thermal hysteresis and coexist with stripe domains, indicating a complex energy landscape within the ferromagnetic phase. These chiral spin textures originate from disorder-induced local DMI, with the interplay between perpendicular magnetic anisotropy and dipolar interactions providing a cooperative mechanism for their stabilization. Together, these results establish Cr$_{1+x}$Te$_2$ as a versatile magnetic compound with highly tunable magnetic phases, providing a compelling platform for investigating skyrmion formation, stability, and unconventional thermodynamic behavior.

\section{Acknowledgments}
The work at the University of Missouri, including sample preparation, XRD, DFT, and ARPES, was primarily supported by the U.S. Department of Energy, Office of Science, Office of Basic Energy Sciences, Division of Materials Science and Engineering, under Grant No. DE-SC0024294. J.H. acknowledges the support from the National Science Foundation DMR-2238254 for single crystal growth. We acknowledge the MonArk NSF Quantum Foundry for magnetic property measurements using MPMS3 SQUID, which is supported by the National Science Foundation (NSF) Q-AMASE-i program under NSF Award No.~DMR-1906383. The STM and magnetization measurements were conducted at the Center for Nanophase Materials Sciences (CNMS), which is a U.S. Department of Energy, Office of Science User Facility. The laser ARPES measurements were supported by the U.S. Department of Energy (DOE), Office of Science, Basic Energy Sciences (BES), Materials Science and Engineering Division.

\section{Methods}
\subsection{Growth of Cr$_{1+x}$Te$_2$ Crystals}
Single crystals were grown by a self-flux method using Te as the flux. Cr and Te powders with a ratio of 1:6 were mixed and vacuum sealed in quartz ampoules. The quartz ampoule was then heated to 1150$^\circ$ C at a heating rate of 50$^\circ$ C/h, kept for 48 hours, slowly cooled to 700$^\circ$ C at a rate of 2$^\circ$ C/h, and centrifuged to separate crystals from excess Te flux.

\subsection{Angle-resolved Photoemission Spectroscopy}
Angle-resolved photoemission spectroscopy measurements were performed in a lab-based system using a Scienta DA30L hemispherical analyzer with a base pressure less than $5 \times 10^{-11}$ mbar and a base temperature of T = 7 K. Samples were illuminated with linearly polarized light using an Oxide h$\nu$ = 11 eV laser system. The light polarization was set perpendicular to the sample and the slits of the detector for standard ARPES measurements. For electronic dispersion measurements, a pass energy of 5 eV and 0.3 mm slit was used for a total energy resolution
= 3.75 meV and momentum resolution = 0.01 ~\AA$^{-1}$.

\subsection{Transmission Electron Microscopy and Lorentz TEM}
TEM specimens were prepared from Cr$_{1+x}$Te$_2$ bulk crystals using a ThermoFisher Helios Hydra plasma focused ion beam (PFIB)/scanning electron microscope (SEM). To use FIB cross-section lift out to prepare a sample in the plan-view geometry, the bulk Cr$_{1+x}$Te$_2$ crystal, which is a layered material, was mounted to a sample stub with the out-of-plane direction oriented perpendicular to the sample stub surface normal. The PFIB cross-section was initially cut using a Xe$^+$ ion beam operated at an accelerating voltage of 30 kV such that the out-of-plane direction of the Cr$_{1+x}$Te$_2$ was parallel to the exposed lamella surface. Following lift-out and attachment to a Cu TEM lift-out grid (Ted Pella), the ion source was switched to Ar$^+$ for subsequent thinning. The specimen was progressively thinned using lower ion beam currents and accelerating voltages, with final thinning performed at 5 kV. Scanning/Transmission Electron Microscope (S/TEM) characterization was carried out using a probe aberration-corrected ThermoFisher Spectra 300 microscope with STEM imaging performed at an accelerating voltage of 300 kV.

The Lorentz TEM (LTEM) lamella sample was prepared using a ThermoFisher Scientific Helio2 focus ion beam (FIB). The cryo Lorentz TEM measurement was performed on a ThermoFisher Scientific Spetra 200 operated at a 200 kV accelerating voltage using a Gatan liquid nitrogen TEM holder.  Images were recorded in a Lorentz mode, in which the main objective lenses were switched off. The defocus length of LTEM images is about -3 mm. Externally out-of-plane magnetic field was generated by applying a small amount of current to the main objective lens.

\subsection{Structural, Transport, and Magnetic Measurements}
Single crystal X-ray diffraction data was collected on a Bruker D8 Venture diffractometer equipped with a Photon II charged pixel area detector using Mo-K$\alpha$ radiation from a microfocus source. A small fragment from an ingot of the Cr$_{1+x}$Te$_2$ was scraped off using a steel blade and mounted on a polyimide sample loop using silicone grease. A hemisphere of unique data was measured at ambient temperature using strategies of 0.5$^\circ$ scans about the omega and phi axes. Unit cell determination, data collection, data reduction, absorption correction and scaling, and space group determination were performed using the Apex4 software suite.  A unit cell search on high intensity peaks (I/$\sigma$ $>$ 10) initially gave the trigonal cell, which was reduced assuming triclinic symmetry. The true monoclinic setting and corresponding transformation matrix were found by comparing the intensity differences for symmetry equivalent reflections (R$_{int}$) for the trigonal cell and all three monoclinic settings. Once the true monoclinic setting was identified, data reduction was repeated with constraints for monoclinic symmetry.

The crystal structure was solved by an iterative dual space method as implemented in SHELXT,  and all atoms were refined anisotropically by full matrix least squares refinement against F$^2$ using SHELXL with an additional parameter for secondary extinctions.  Partial occupancy chromium sites were identified by their large thermal parameters or by the presence of significant difference map peaks too small to be full occupancy atoms, and for these sites the occupancy parameters were refined freely. Olex2 was used as a graphical interface for model building and structure visualization.

Samples were cleaved under ultrahigh vacuum conditions and measured using a SCIENTA OMICRON LT Nanoprobe STM at 4.6 K. STM topographic images were acquired in constant current mode using platinum iridium tips. Magnetic measurements were performed using a Quantum Design Magnetic Property Measurement System (MPMS3) at the Center for Nanophase Materials Science (CNMS) at Oak Ridge National Laboratory. The field cooled (FC) and zero field cooled (ZFC) measurements were carried out between the temperature range of 2--300 K. Magnetic hysteresis measurements were taken with a maximum sweeping magnetic field of 70000 Oe. ac measurement were taken at ac field of 3 Oe at a frequency of 100 Hz and 0 Oe DC field.  
\subsection{First-Principle Calculations}
First-principles calculations with DFT were performed by using the Vienna ab Initio Simulation Package (VASP). We used the Perdew–Burke–Ernzerhof (PBE) form of the exchange correlation functional. All the calculations were performed with a plane-wave cut-off energy of 400 eV on the 20 × 20 × 1 Monkhorst-Pack k-point mesh. A unit cell was constructed using the measured lattice constants from STM and the atoms were allowed to fully relax. Band structures were calculated with the inclusion of SOC.

\bibliographystyle{apsrev4-2}
\bibliography{Cr5Te8}

\begin{figure}
\includegraphics[width=1.0\linewidth]{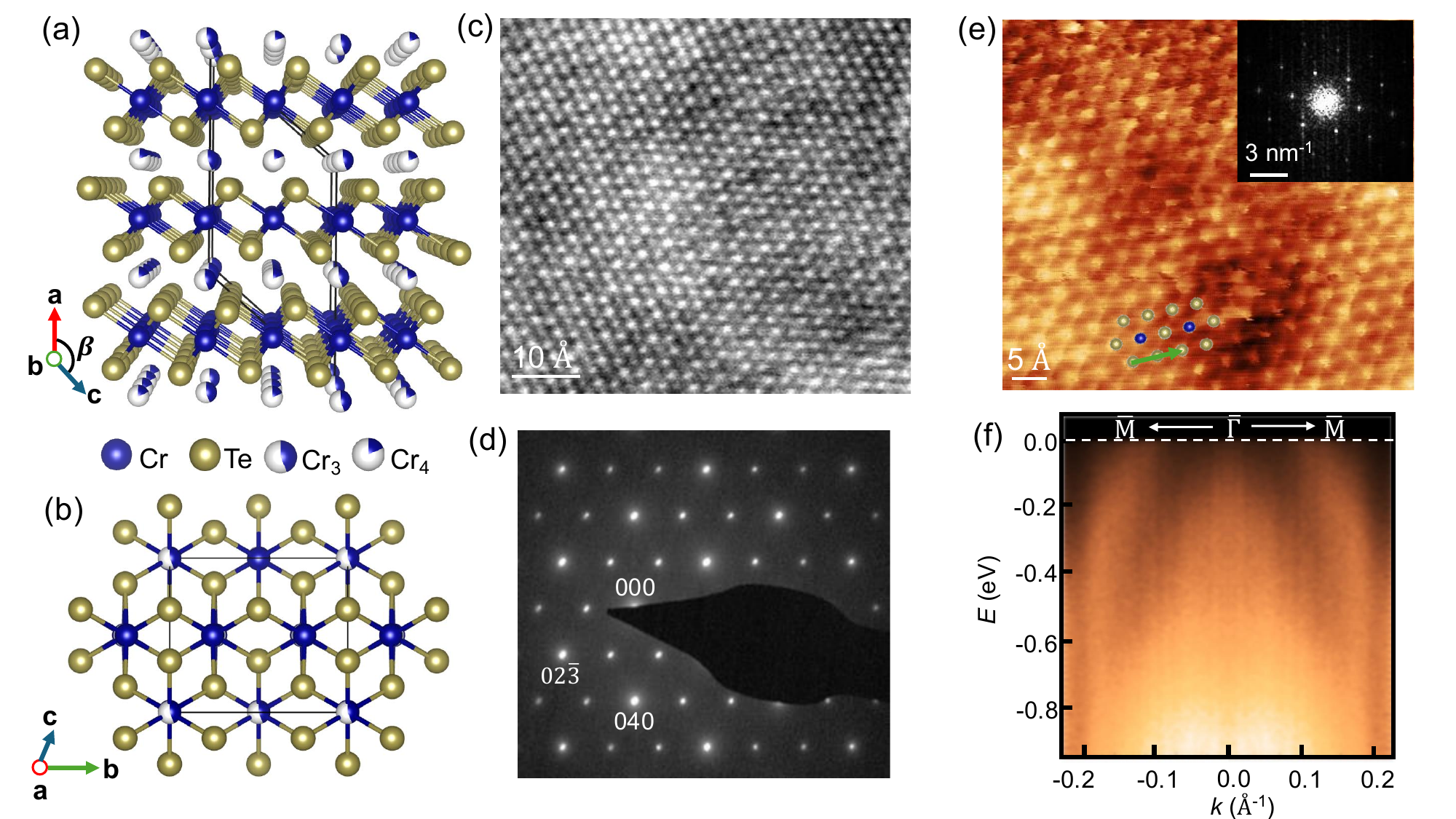}
\caption{Structural and Electronic Properties of Cr$_{1+x}$Te$_2$. (a) Side view along the in-plane (010) direction and (b) top view along the out-of-plane (100) direction of the crystal structure as measured by SCXRD. The partially occupied Cr sites, Cr$_3$ and Cr$_4$, can be seen within the vdW gaps of the CrTe$_2$ layers with a 2$\times$2 periodicity. (c) HAADF STEM image taken along the OOP direction. (d) SAED pattern with diffraction spots indexed according to the SCXRD result. (e) Atomic-resolution STM image of Cr$_{1+x}$Te$_2$ surface. The green arrow marks the primitive vector $\mathbf{b}$. The FFT of the STM image is shown in the inset. (f) ARPES results taken along M-$\Gamma$-M direction. The observed hole pocket confirms the metallic behavior of the sample.}%
\end{figure}

\newpage

\begin{figure}
\includegraphics[width=1.0\linewidth]{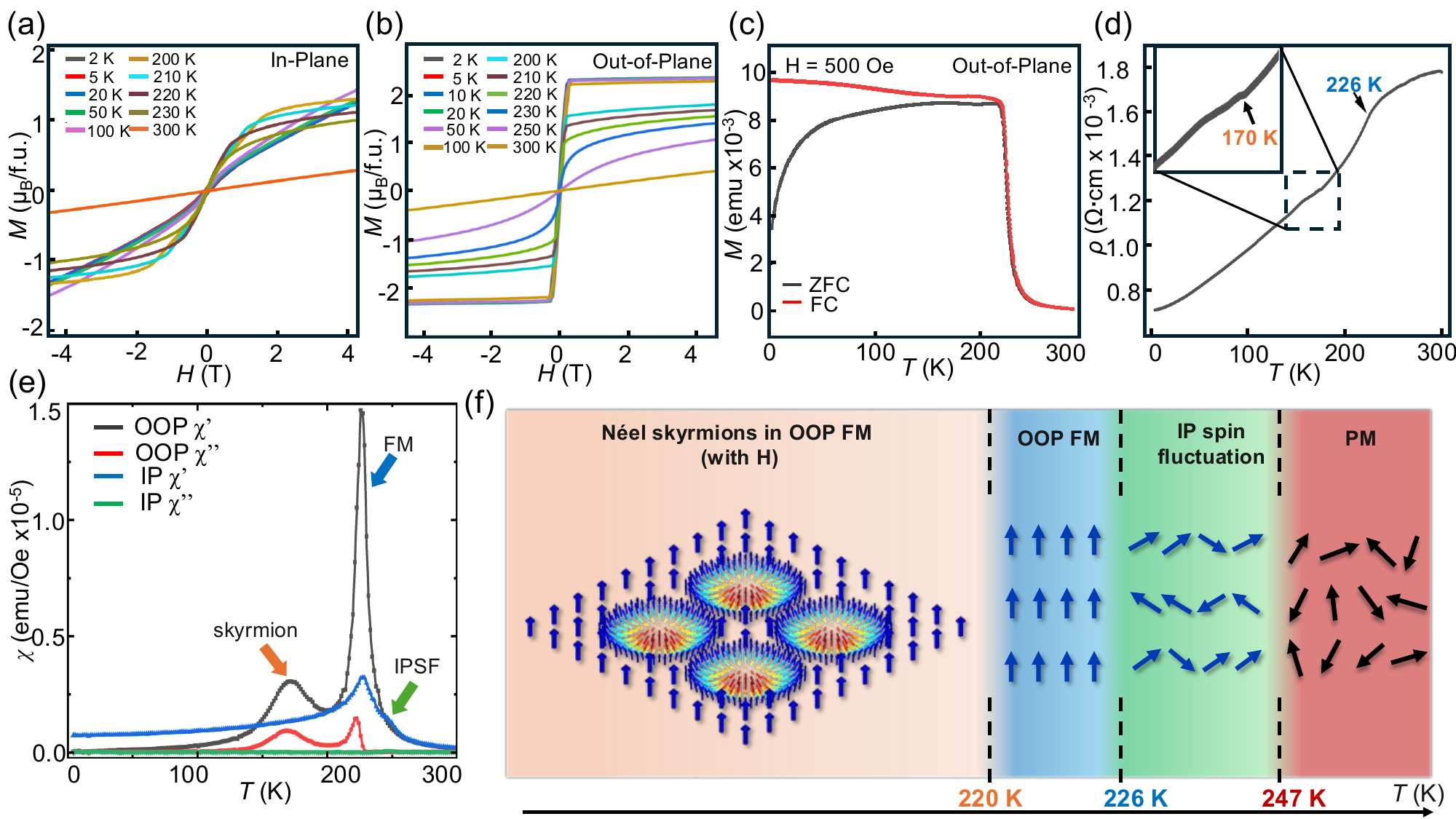}
\caption{Magnetic and Transport Measurements of Cr$_{1+x}$Te$_2$. (a,b) Field-dependent magnetization measurements in the (a) IP  and (b) OOP field configurations. The OOP magnetization fully saturates, whereas the IP magnetization does not, indicating an out-of-plane magnetic easy axis in the system. (c) $M$–$T$ curves measured with an OOP field of 500~Oe. A clear ferromagnetic transition is observed at 226~K. The plot includes both ZFC and FC measurements. (d) Measured longitudinal resistivity as a function of temperature for Cr$_{1+x}$Te$_2$. A slope change occurs at the Curie temperature ($T_\mathrm{C}$) of 226~K, confirming the ferromagnetic ordering. An additional kink-like feature at 170~K is highlighted in the inset. (e) AC magnetic susceptibility measured at a frequency of 100 Hz for both OOP and IP configurations. Four distinct magnetic phases can be observed in this plot. (f) Comprehensive magnetic phase diagram of the Cr$_{1+x}$Te$_2$ system. A field-induced Néel skyrmion bubble phase coexists with the ferromagnetic background below 220~K. Above this temperature the skyrmion phase disappears, leaving a homogeneous out-of-plane ferromagnetic state up to $T_\mathrm{C}=226$ K, followed by an in-plane spin fluctuation regime and the paramagnetic state at higher temperature.} 
\end{figure}

\newpage
\begin{figure}
\includegraphics[width=1.0\linewidth]{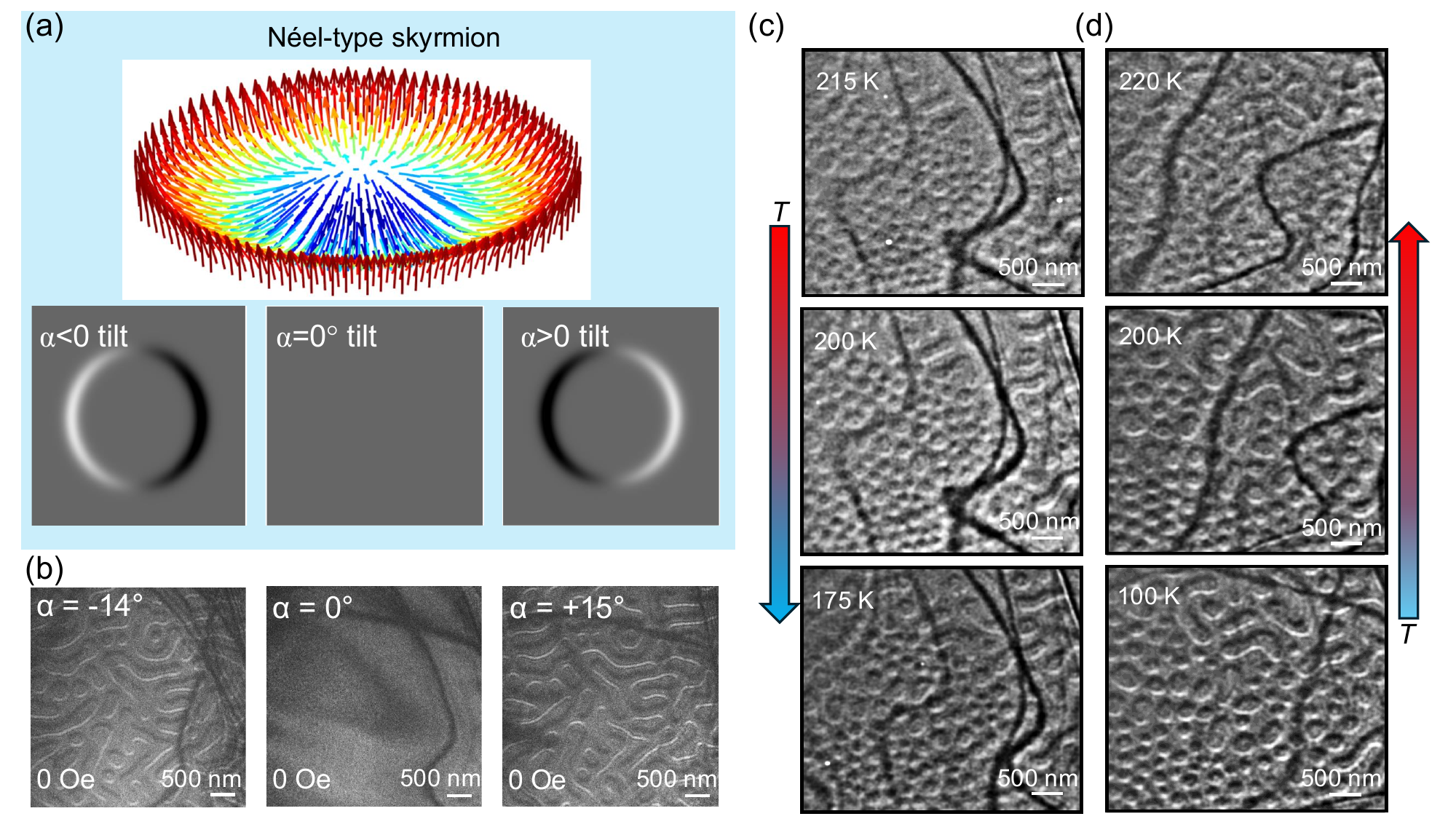}
\caption{Néel-type skyrmion and Lorentz TEM results of Cr$_{1+x}$Te$_2$. (a) Schematic of the magnetic spin texture of a Néel-type skyrmion (top) and simulated tilt-dependent LTEM image of a Néel-type skyrmion at different tilts (bottom). No contrast is seen at zero tilt. Upon applying a tilt, contrast emerges and reverses when the tilt changes from negative to positive. (b) LTEM results of Cr$_{1+x}$Te$_2$ measured with zero applied field at 120~K. Stripe domains only appear when tilt is applied and undergo a contrast reversal from negative to positive. This indicates Néel-type domain walls as the magnetic ground state. (c) Néel-type skyrmions appear in Cr$_{1+x}$Te$_2$, upon cooling, near the $T_\mathrm{C}$ with a field of 300~Oe applied. The size of skyrmions increases as the temperature is lowered. (d) LTEM images of skyrmions as the sample is warmed up under the same 300~Oe applied field. The skyrmion size remains unchanged from 100~K to 200~K during this warming process.}%
\end{figure}

\newpage
\begin{figure}
\includegraphics[width=1.0\linewidth]{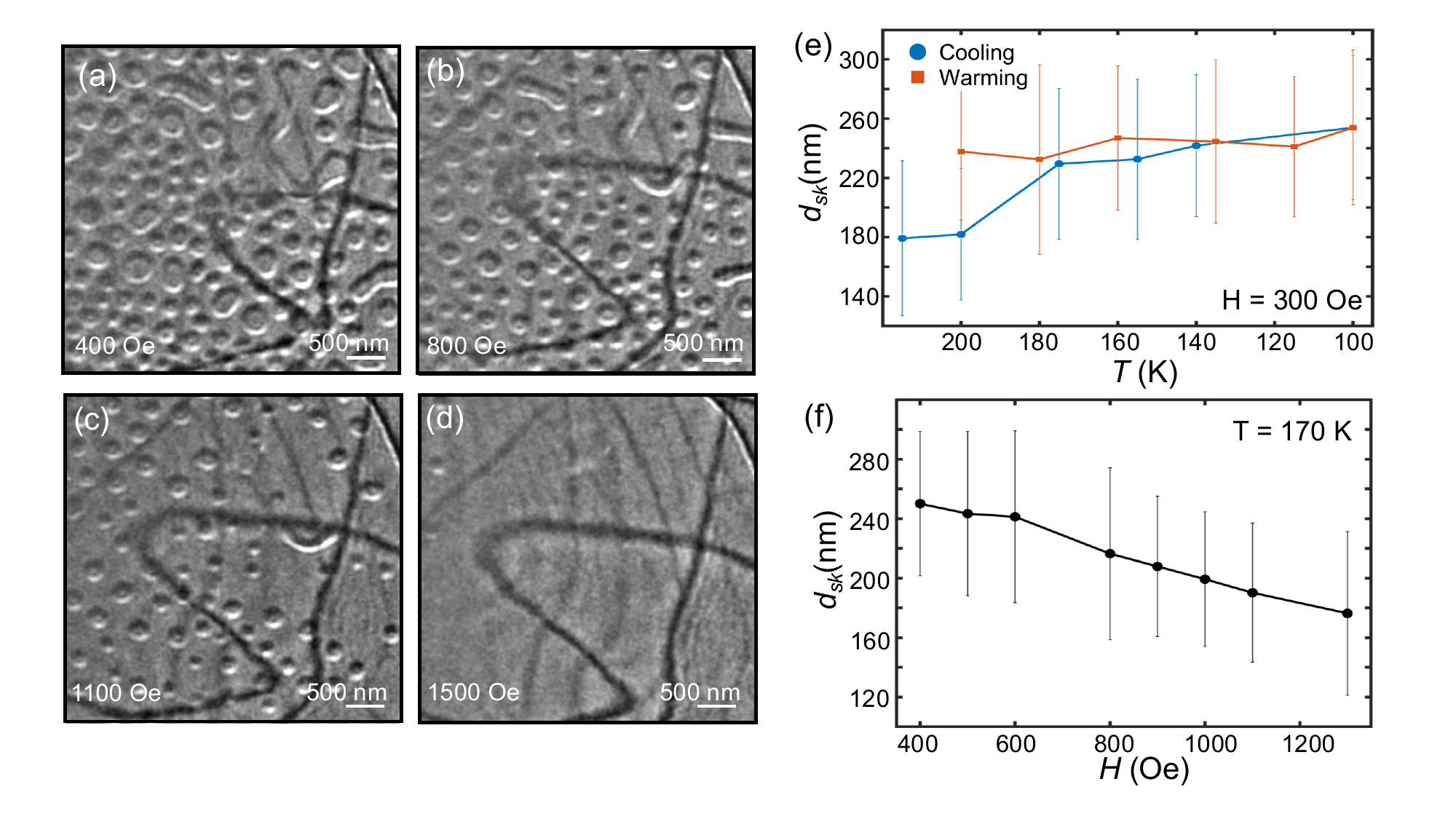}
\caption{Temperature and field dependence of skyrmion size. (a-d) LTEM images of Cr$_{1+x}$Te$_2$ taken at 170~K under various applied field strengths. (e) Skyrmion diameter (d$_{sk}$) measured with a field of 300 Oe applied as a function of temperature, showing both cooling and warming behavior. (f) Measured skyrmion diameter as a function of the applied magnetic field strength.} 
\end{figure}

\end{document}